\documentclass[11pt]{article}
\usepackage{citesort,amsmath,xspace,graphicx,amssymb}
\usepackage{mathptmx,helvet,courier}
\usepackage{setspace}
\usepackage{paralist}

\def\hlinefill{\leaders\hrule height3pt depth-2.5pt\hfill}
\def\emrule{\thinspace\hbox to .75em{\hlinefill}\thinspace}
\makeatletter
\def\@cite#1#2{({#1\if@tempswa : #2\fi})}
\def\@biblabel#1{}
\def\@citex[#1]#2{%
  \let\@citea\@empty
  \@cite{\@for\@citeb:=#2\do
    {\@citea\def\@citea{;\penalty\@m\ }%
     \edef\@citeb{\expandafter\@firstofone\@citeb\@empty}%
     \if@filesw\immediate\write\@auxout{\string\citation{\@citeb}}\fi
     \@ifundefined{b@\@citeb}{\mbox{\reset@font\bfseries ?}%
       \G@refundefinedtrue
       \@latex@warning
         {Citation `\@citeb' on page \thepage \space undefined}}%
       {\hbox{\csname b@\@citeb\endcsname}}}}{#1}}

\def\thebibliography#1{\section*{References}\list
 {\arabic{enumi}.}{
 \leftmargin0pt%
 \itemsep=2pt plus 1pt\parsep=0pt
 \usecounter{enumi}}
 \def\newblock{\hskip .11em plus .33em minus -.07em}
 \sloppy
 \sfcode`\.=1000\relax}

\def\@oddhead{\hbox{\textit{Johnson}} \hfil
\hbox{\textit{Dependent Poisson Processes}}}
\def\@oddfoot{\hbox{} \hfil \thepage \hfil \hbox{\today}}
\def\fnum@figure{\textbf{Figure \thefigure}}
\def\fnum@table{\textbf{Table \thetable}}

\long\def\@makecaption#1#2{%
  \vskip\abovecaptionskip
  \advance\baselineskip -2pt
  \sbox\@tempboxa{#1: #2}%
  \ifdim \wd\@tempboxa >\hsize
    {\small #1: #2}\par
  \else
    \global \@minipagefalse
    \hbox to\hsize{\hfil\box\@tempboxa\hfil}%
  \fi
  \vskip\belowcaptionskip}
\renewcommand{\section}{\@startsection {section}{1}{\z@}%
                                   {-.75ex \@plus -1ex \@minus -.2ex}%
                                   {.75ex \@plus.2ex}%
                                   {\reset@font\large\bfseries}}
\renewcommand{\subsection}{\@startsection{subsection}{2}{\z@}%
                                     {-.5ex\@plus -.4ex \@minus -.2ex}%
                                     {.5ex \@plus .2ex}%
                                     {\reset@font\normalsize\bfseries}}
\renewcommand{\subsubsection}{\@startsection{subsubsection}{3}{\z@}%
                                     {-1ex\@plus -.5ex \@minus -.2ex}%
                                     {1.5ex \@plus .2ex}%
                                     {\reset@font\normalsize\bfseries}}
\renewcommand{\paragraph}{\@startsection{paragraph}{4}{\z@}%
                                    {.5ex \@plus1ex \@minus.2ex}%
                                    {-.5em}%
                                    {\reset@font\normalsize\bfseries}}
\renewcommand{\subparagraph}{\@startsection{subparagraph}{4}{\parindent}%
                                       {1ex \@plus1ex \@minus 2ex}%
                                       {-1em}%
                                      {\reset@font\normalsize\bfseries}}

\def\@sect#1#2#3#4#5#6[#7]#8{\ifnum #2>\c@secnumdepth
     \def\@svsec{}\else
     \refstepcounter{#1}\edef\@svsec{\csname the#1\endcsname\hskip 1em }\fi
     \@tempskipa #5\relax
      \ifdim \@tempskipa>\z@
        \begingroup #6\relax
          \@hangfrom{\hskip #3\relax\@svsec}{\interlinepenalty \@M #8\par}%
        \endgroup
       \csname #1mark\endcsname{#7}\addcontentsline
         {toc}{#1}{\ifnum #2>\c@secnumdepth \else
                      \protect\numberline{\csname the#1\endcsname.}\fi
                    #7}\else
        \def\@svsechd{#6\hskip #3\@svsec #8\csname #1mark\endcsname
                      {#7}\addcontentsline
                           {toc}{#1}{\ifnum #2>\c@secnumdepth \else
                             \protect\numberline{\csname the#1\endcsname}\fi
                       #7}}\fi
     \@xsect{#5}}
\newdimen\fboxrules
\newdimen\fboxrulel
\long\def\eqnbox#1{\leavevmode\setbox\@tempboxa\hbox{$\displaystyle #1$}\@tempdima\fboxrules
    \advance\@tempdima \fboxsep \advance\@tempdima \dp\@tempboxa
   \hbox{\lower \@tempdima\hbox
  {\vbox{\hrule \@height \fboxrules
          \hbox{\vrule \@width \fboxrules \hskip\fboxsep
          \vbox{\vskip\fboxsep \box\@tempboxa\vskip\fboxsep}\hskip 
                 \fboxsep\vrule \@width \fboxrulel}
                 \hrule \@height \fboxrulel}}}}
\makeatother

\newcommand{\pdf}{p}

\newcommand{\mgf}[1]{M_{#1}(s)}
\newcommand{\step}[1]{\mathrm{u}(#1)}

\newcommand{\E}{\mathsf{E}}

\newcommand{\ratesym}{\lambda}

\newcommand{\ppcount}[1]{N_{#1}}

\newcommand{\isi}{\tau}

\begin{document}
\begin{center}
\textbf{\large The Correlation Function of Multiple Dependent Poisson Processes Generated by the Alternating Renewal Process Method}\\
\smallskip
\textit{Don H. Johnson}\\
Computer and Information Technology Institute\\
Electrical \& Computer Engineering Department, MS380\\
Rice University \\
Houston, Texas 77005--1892\\
\textit{dhj}@rice.edu\\[6pt]
\end{center}
\sloppy

\begin{abstract}
\noindent
We derive conditions under which alternating renewal processes can be used to construct correlated Poisson processes.
The pairwise correlation function is also derived, showing that the resulting correlations can be negative.
The technique and the analysis can be extended to the generation of two or more dependent renewal processes.
\end{abstract}

\doublespacing
\section{Introduction}
\begin{figure}
\centerline{\includegraphics{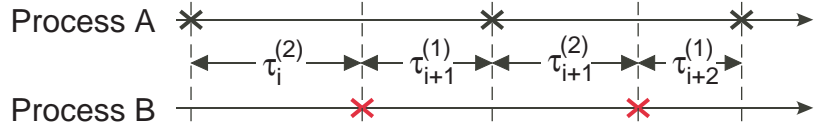}}
\caption[]{\label{fig:alt}
An alternating renewal process consists of a sequence of interval pairs $\isi_i^{(1)},\isi_i^{(2)}$.
Each member of the pair and the pairs are statistically independent stochastic quantities.
Dependent renewal processes are derived by assigning intervals ending in an interval drawn from interval distribution~1 to one process (here process~A) and those ending in an interval drawn from distribution~2 to the other.
}
\end{figure}

Bruce Knight first suggested using specially chosen alternating renewal processes to construct correlated Poisson processes.
An alternating renewal process has successive intervals drawn in turn from one of two probability distributions.
For example, interval $\isi_i^{(1)}$ is drawn from $\pdf_1(\isi)$ and $\isi_i^{(2)}$ from $\pdf_2(\isi)$ independently of the first.
Process construction continues in this fashion.
To derive dependent renewal processes, events in the alternating renewal process are assigned to one or the other, with events ending intervals drawn from process~1 assigned to one (call it process~A) and events ending in process~2 intervals assigned to the other (process~B) (Figure~\ref{fig:alt}).
Note that the sum of successive intervals, no matter which pair is chosen, has the probability distribution $\pdf(\isi)$ given by $\pdf_1(\isi)\star\pdf_2(\isi)$, where $\star$ denotes convolution.
Thus, the result if Knight's construction is, in general, two statistically dependent, identically distributed renewal processes.
To make each member of the pair be Poisson, $\pdf(\isi)$ must be exponential:
$\pdf(\isi)=\ratesym e^{-\ratesym\isi}\step{\isi}$.

\section{Analysis of the Method}
The requirements for the component interval distributions to achieve dependent Poisson (dependent renewal as well) are best expressed in the frequency domain using moment-generating functions.
Defining
\[
\mgf{\isi}=\int_{0-}^{\infty}\!\pdf(\isi)e^{s \isi}\,d\isi\;,
\]
we have $\mgf{\isi}=\mgf{1}\cdot\mgf{2}$.
Thus, to construct dependent renewal processes, we need only chose the desired interval distribution and find two interval distributions that satisfy this constraint.
The tricky part is not satisfying the constraint, but rather insuring that the two components correspond to interval distributions (i.e., they cannot be negative).

For dependent Poisson processes, each of which must have $\mgf{\isi}=1/(1-s/\ratesym)$, Knight's specific suggestion is the choice
\[
\mgf{1} = \frac{1}{(1-s/\ratesym)^{g_1}}
\quad
\mgf{2} = \frac{1}{(1-s/\ratesym)^{g_2}}\;.
\]
Each is a viable interval distribution, and to produce dependent Poisson processes, we only need $g_1+g_2=1$.
Daniel Fisher's suggestion is more complicated.
\[
\mgf{1} = \frac{g^2}{a^2} \left(\frac{a-s}{g-s}\right)^2 \frac{1}{1-s}
\quad
\mgf{2} = \frac{a^2}{g^2} \left(\frac{g-s}{a-s}\right)^2
\]
Clearly, the product of the two is $1/(1-s)$ and does generate Poisson processes like Knight's procedure having an average rate of one.
As we will show, this construction is overly complicated.

\section{Correlation Function}
Quantifying the dependence between the generated processes can be expressed by the cross-correlation function defined as
\[
R_{A,B}(\isi)\equiv \E\left[\frac{d \ppcount{A,t}}{dt}\cdot\frac{d \ppcount{B,t+\isi}}{dt}\right]\, ,\isi>0\,,
\]
where $\ppcount{t}$ is the counting process, defined as the number of events that has occurred up to time $t$.
The derivative of the counting process creates impulses at each event occurrence.
The cross-correlation can be explicitly calculated for any $\mgf{1}$, $\mgf{2}$.
To find $R_{A,B}(\isi)$ for $\isi<0$, we evaluate $R_{B,A}(\isi)$ for $\isi>0$.
The expected value is interpreted as the probability of an event occurring in process~B at time $t+\isi$ given that an event occurred in process~A at time $t$ multiplied by the probability of an event occurring in the first process at time $t$.
Expressing the cross-correlation function as a function of $\isi$ only means that the processes are jointly stationary.

As an example of how the covariance function can be computed, consider the auto-correlation function of a renewal process.
The required conditional probability equals the sum of the probabilities that $k$ events occur between times $t$ and $t+\isi$ when an event occurs at time $t+\isi$.
For $k=0$, the conditional probability equals the interval distribution.
For $k=1$, it equals the probability that two intervals sum to $\isi$.
For any $k$, the conditional probability equals the probability $k+1$ intervals sum to $\isi$.
These probabilities amount to the $k$-fold convolution of the interval distribution with itself, which is most easily calculated using the moment generating function.
\[
\pdf(\isi)\underbrace{\star\dots\star}_{k-1} \pdf(\isi) \longleftrightarrow [\mgf{\isi}]^k
\]
Therefore,
\[
R_{A,A}(\isi)\longleftrightarrow \bar{\ratesym}\sum_{k=1}^{\infty} [\mgf{\isi}]^k = \bar{\ratesym}\frac{\mgf{\isi}}{1-\mgf{\isi}}
\]
Here, $\bar{\ratesym}$ is the average rate at which events occur.
For a renewal process, $\bar{\ratesym}=1/\E[\isi]$.
As a check, consider the Poisson process, in which $\mgf{\isi}=\ratesym/(\ratesym-s)$.
In this case, $\bar{\ratesym}=\ratesym$ and the transform of the correlation function equals
\[
R_{A,A}(\isi)\longleftrightarrow \ratesym \frac{\frac{\ratesym}{\ratesym-s}}{1-\frac{\ratesym}{\ratesym-s}} =-\frac{\ratesym^2}{s} \;.
\]
The inverse transform of this quantity is the unit-step: $R_{A,A}(\isi)=\ratesym^2\step{\isi}$.
By subtracting the square of the mean event rate we obtain the covariance function, equal to zero in this case.
This result expresses the fact that a Poisson process is white noise.
The general expression for the autocovariance function is
\[
K_{A,A}(\isi)\longleftrightarrow \bar{\ratesym}\frac{\mgf{\isi}}{1-\mgf{\isi}}+\frac{\bar{\ratesym}^2}{s}
\]

Returning to the cross-correlation function, we need the conditional probability of an event occurring in process~B at time $t+\isi$ given that an event occurs in~A at time $t$.
Because the alternating renewal method creates Poisson processes from renewal processes, we can use the moment-generating function technique here as well.
The first term ($k=0$) corresponds to the interval distribution between an event in the reference process and the next event in the other.
Because of the construction method, this interval is determined by one of the component process's interval distribution, $\isi_i^{(2)}$.
The next term in the cross-correlation function corresponds to the convolution of this interval distribution with the interval distribution of the derived Poisson process.
The third term corresponds to the convolution of the second process's interval distribution with convolution of the Poisson process's interval distribution with itself.
Thus,
\begin{align*}
R_{A,B}(\isi)\longleftrightarrow \bar{\ratesym}\sum_{k=0} \mgf{2} [\mgf{\isi}]^k &= \bar{\ratesym}\left[\mgf{2} + \mgf{2}\frac{\mgf{\isi}}{1-\mgf{\isi}}\right] \\
&= \ratesym \mgf{2} - \frac{\ratesym^2}{s} \mgf{2} \;,
\end{align*}
the latter equation being the only one employing the Poisson assumption.
Adding $\ratesym^2/s$, we obtain the Laplace transform of the cross-covariance function.
\[
\eqnbox{K_{A,B}(\isi)\longleftrightarrow \ratesym \mgf{2} - \frac{\ratesym^2}{s}\bigl(\mgf{2}-1\bigr)}
\]
Once inverse-transformed, we can judge whether the two processes are positively or negatively dependent.
Note that to obtain the expression for negative $\isi$, we simply replace $\mgf{2}$ by $\mgf{1}$.

For Knight's example, the covariance function equals
\[
\frac{\ratesym^{g+1}\isi^{g-1}  e^{-\ratesym\isi }-\ratesym^2 
    \Gamma (g,\ratesym  \tau )}{\Gamma (g)}\,.
\]
Here, $\Gamma(\cdot)$ is the Gamma function and $\Gamma(\cdot,\cdot)$ is the incomplete Gamma function.
Plotting this quantity for several values of $g$ and $\ratesym$ indeed confirms his statement that the two Poisson processes are positively correlated.

For Fisher's example, the covariance function equals
\[
K_{A,B}(\isi)=
  \begin{cases}
    \frac{g-a}{g^2} e^{-a\isi} \bigl(2a^2-(g+a)+a(a-1)(g-a)\isi\bigr)+\frac{a^2}{g^2}\delta(\isi), & \isi\ge0\\
   \frac{g-a}{a^2} e^{+g\isi} \bigl(g+a+g(g-a)\isi\bigr), & \isi<0
  \end{cases}
\]
Note that his construction has $\ratesym=1$, which means the exact role of $\ratesym$ is hidden.
In any case, this expression indicates that negative correlations can occur for many choices of $a,g$.
For $\isi>0$, negative correlations can occur for $\isi$ less than $\frac{a+g-2a^2}{a(a-1)(g-a)}$;
for $\isi<0$, they can occur for $\isi$ less than $-\frac{g+a}{g(g-a)}$.
Consequently, the negative correlations can occur only for lags $\isi$ asymmetrically located about the origin.

A whole host of examples follow from this framework.
We know that we must have
\[
\mgf{2}=\frac{\ratesym}{\ratesym-s}\cdot\frac{1}{\mgf{1}} \;.
\]
For each suggested $\mgf{1}$, we must check the range of parameter values over which $\mgf{2}$ is a valid moment generating function:
its inverse transform must be non-negative.
Now armed with a valid alternating renewal process, we can investigate the dependence structure of the derived Poisson processes.

\paragraph{Example.}
If $\mgf{1}=a/(a-s)$, we must have $a>\ratesym$ for $\pdf_2(\isi)>0$.
In this case,
\[
K_{A,B}(\isi)=\ratesym(a-\ratesym)e^{-a\isi}\step{\isi} + \frac{\ratesym^2}{a}\delta(\isi) \;.
\]
Thus, the two Poisson processes are positively correlated.

\paragraph{Example.}
More interesting is the example
\[
\mgf{1}=\frac{a_2}{a_1}\cdot \frac{a_1-s}{a_2-s} \;.
\]
The corresponding interval distribution is
\[
\pdf_1(\isi)=\frac{a_2}{a_1}\left[(a_1-a_2)e^{-a_2\isi}\step{\isi} + \delta(\isi)\right]\;,
\]
which demands that $a_1>a_2$.
The interval distribution corresponding to $\mgf{2}$ is
\[
\pdf_2(\isi)=\frac{a_1\ratesym}{a_2(a_1-\ratesym)} \left[(a_1-a_2)e^{-a_1\isi} + (\ratesym-a_2)e^{-\ratesym\isi}\right]\step{\isi}\;,
\]
which requires $a_1>\ratesym>a_2$ to be positive.
We obtain
\[
K_{A,B}(\isi)= \begin{cases}
-\frac{\ratesym(a_1-a_2)(\ratesym-a_2)}{a_1} e^{-a_2\isi} + \frac{a_2\ratesym}{a_1}\delta(\isi), & \isi\ge0\\
\frac{\ratesym^2(a_1-a_2)}{a_2}e^{a_1\isi}, & \isi<0
\end{cases}
\]
Here, negative correlation occurs for positive lags, positive correlation for negative lags.
Also note the presence of the impulse in the correlation function, a direct consequence of the impulse in process~1's interval distribution.

\section{Summary}
These results provide an analytic prediction of the correlation between Poisson processes constructed from alternating renewal processes.
The technique is not restricted to Poisson processes;
correlated renewal process can be generated as well.
The two component renewal processes must be chosen so that the convolution of their interval distributions equal the desired one.
Furthermore, more than pairs of dependent renewal processes can be generated this way and analytic expressions for the correlation functions derived by simply extending the approach described here.

One limitation of this method is how to construct dependent \emph{non-stationary} Poisson processes.
Doing so requires the underlying alternating renewal process to be time-varying.
It is difficult to construct such processes and even more difficult to analyze the cross-correlation function between them.

From a formal viewpoint, the resulting Poisson processes constructed from an alternating renewal process are \emph{not} what a probabilist would term jointly Poisson.
Analogous to Gaussian random variables that are jointly Gaussian, jointly Poisson processes have special analytic properties.
The defining characteristic of jointly distributed random variables that enjoy special status as limiting distributions\emrule the Central Limit Theorem in the Gaussian case and the superposition of point processes converge in the limit to a Poisson process\emrule is \emph{infinite divisibility}.
This concept requires that the quantity in question always be expressed as a sum of an arbitrary number of constituents that have the same distributional form (\textit{i.e.}, they differ only in parameter values).
It has been shown that two jointly defined Poisson processes $\ppcount{A,t}$ and $\ppcount{B,t}$ that are infinitely divisible can always be constructed by
\begin{align*}
\ppcount{A,t} &= \ppcount{1,t}+\ppcount{t}^0\\
\ppcount{B,t} &= \ppcount{2,t}+\ppcount{t}^0\;,
\end{align*}
where $\ppcount{1,t}$ and $\ppcount{2,t}$ are statistically independent Poisson processes and $\ppcount{t}^0$ is another Poisson process statistically independent of the others but shared in common between the constructed processes.
Consequently, the correlations between the processes $\ppcount{A,t}$ and $\ppcount{B,t}$ occur because of the common Poisson process component, which means simultaneous occurrence of events produces the correlations.
These dependencies thus have two important properties:
\begin{inparaenum}[(1)]
\item
Correlations \emph{must} be non-negative and
\item
occur simultaneously, meaning the two process's cross-correlation function has no temporal extent.
\end{inparaenum}
Consequently, jointly Poisson processes share only one property with those created from an alternating renewal process:
the marginal (individual) point processes are Poisson.
Although this construction method might prove useful, analytically dealing with it or its variants will be difficult.
Generating non-Poisson renewal processes this way may be more fruitful, but the theory of jointly defined renewal processes is undeveloped.

\clearpage
\singlespacing
\bibliographystyle{apalike}
\bibliography{spikes1,spikes2}

\end{document}